\documentclass[]{aa}  

\usepackage{natbib}
\bibpunct{(}{)}{;}{a}{}{,}
\usepackage{graphicx}
\usepackage{revsymb}	
\usepackage{amsmath}	
\usepackage[usenames]{color}	
\usepackage{epstopdf}	
\usepackage{txfonts}
\usepackage{amssymb}
\usepackage{color}
\usepackage[justification=centering]{caption}
\usepackage{geometry} 
\usepackage[parfill]{parskip} 
\usepackage{amssymb}
\usepackage{slantsc}
\usepackage{array}
\usepackage{url}
\usepackage{amsfonts}
\usepackage[colorlinks=true,linkcolor=blue, urlcolor=blue, citecolor=blue]{hyperref}
\usepackage{sidecap}
\usepackage{graphicx}
\usepackage{xcolor}
\usepackage{nicefrac}
\usepackage{subfigure}
\usepackage{epsfig}
\colorlet{rouge}{red!70!darkgray}

\begin{document}
\title{A data-driven estimate of the protosolar helium mass fraction}
\author{G. Buldgen\inst{1} \and M. Kunitomo\inst{2} \and A. Noels\inst{1} \and T. Guillot\inst{3} \and R. Scuflaire\inst{1} \and N. Grevesse \inst{1,4}}
\institute{STAR Institute, Université de Liège, Liège, Belgium \and Department of Physics, Kurume University, 67 Asahimachi, Kurume, Fukuoka 830-0011, Japan \and Universit\'e C\^ote d'Azur, Observatoire de la C\^ote d'Azur, CNRS, Laboratoire Lagrange, Bd de l'Observatoire, CS 34229, 06304 Nice cedex 4, France \and Centre Spatial de Liège, Université de Liège, Angleur-Liège, Belgium}
\date{January, 2026}
\abstract{The protosolar helium mass-fraction is a fundamental ingredient of solar, planetary models and enrichment laws used to model stellar populations. However, the assumed values often relies on simplifying descriptions of the transport of chemicals in solar models. Furthermore, they are based on the inferred helium mass fraction in the solar convective envelope, which is itself sensitive to uncertainties in the equation of state of the solar material.}
{We aim at updating the reference protosolar helium abundance by taking into account the effects of macroscopic mixing at the base of the convective zone and using more recent determinations of the helium mass fraction in the convective envelope.}
{We combine results from our own inversions of the composition of the solar envelope to spectroscopic abundances, as well as values in the literature to provide a robust interval of the current helium mass fraction in the convective zone. We combine this measurement to solar models taking into account light element depletion to provide an udpated protosolar helium abundance.}
{We show that macroscopic mixing at the base of the convective envelope of the Sun cannot be neglected to infer the protosolar helium abundance. We demonstrate that as soon as this effect is included, the protosolar helium abundance is significantly reduced and that lithium and beryllium depletion can be used to calibrate this effect over the solar evolution. We find a revised interval of primordial helium mass fraction of $0.27575\pm0.00315$ slightly lower than previous estimates when combining our latest estimate of surface helium mass fraction and spectroscopic abundances. We find that the effects of macroscopic mixing are partially compensated by an increase in the inferred solar helium mass fraction in recent studies but also derive more precise estimates based on various reference works in the litterature. If the usual surface helium mass fraction is used, the primordial helium mass fraction drops to $0.2669\pm 0.00415$ as a result of the inclusion of macroscopic mixing. The dominant source of uncertainty on this value is found to be the surface helium abundance inferred from helioseismic constraints and more specifically, the impact on the equation of state of the solar material on this inference result.}{}
\titlerunning{Data-driven estimate of the protosolar helium mass fraction} \authorrunning{G. Buldgen et al.}
\maketitle
\section{Introduction}

The protosolar helium abundance, denoted $\rm{Y_{P}}$, is an important ingredient of planetary models, particularly when simulating the evolution of the giant planets of the Solar system \citep[see e.g.][]{Guillot1997,Nettelmann2015,Mankovich2016,Howard2024,Nettelmann2024,Nettelmann2025}. Thanks to helioseismology, we are able to estimate the current helium mass fraction in the convective zone of the Sun, denoted $\rm{Y_{CZ}}$ \citep{Vorontsov1991,BasuYSun,RichardY,DiMauro2002,BasuY2004,Vorontsov13,VorontsovSolarEnv2014,Buldgen2024Z}. Each of these studies find a slightly different (though often overlapping at $1\sigma$) value. The most commonly cited is $\rm{Y_{CZ}}=0.24875 \pm 0.0035$ from \citet{BasuY2004}, which is compatible to the value reported by \citet{RichardY} of $\rm{Y_{CZ}}=0.2480 \pm 0.0020 $. However, these results remain bound to the accuracy of the equation of state of the solar material and in practice, the reported values and associated uncertainties may significantly differ (see Sect. \ref{Sec:EstimateHe}) as can be seen from the results from \citet{DiMauro2002} $\rm{Y_{CZ}}=0.2539 \pm 0.0005 $ when using the OPAL equation of state \citep{OPALEOSI, OPALEOSII} and $\rm{Y_{CZ}}=0.2457 \pm 0.0005$ when using the MHD equation of state \citep{MHDI, MHDII, MHDIII, MHDIV}. Later studies by \citet{Vorontsov13} and \citet{VorontsovSolarEnv2014}, combining both OPAL and SAHA-S \citep{Gryaznov2004, Gryaznov2006, Gryaznov2013, Baturin2013} equations of state, report a lower precision $\rm{Y_{CZ}}=0.2475 \pm 0.0075 $ and $\rm{Y_{CZ}}=0.2525 \pm 0.0075 $, dominated by uncertainties in the equation of state.

The main source of discrepancies between all these studies is the equation of state of the solar plasma, thus motivating continued improvements of the equation of state used in solar and stellar models to achieve higher precision \citep{Antia1994,Baturin2003,Trampedach2006,Baturin2025,Trampedach2025}. All the results above are very consistent with each other, but their respective precision vary significantly, with more recent estimate by \citet{Vorontsov13} and \citet{VorontsovSolarEnv2014} being less precise by a factor $2$ over the usual value taken from \citet{BasuY2004}. As we will see below, such a large uncertainty has a direct impact on the protosolar helium mass fraction. To infer the protosolar helium mass fraction, it is necessary to link the current surface helium mass fraction to the protosolar one by assuming/simulating the evolution of helium during the life of the Sun. The best way to do this is by computing solar models and analysing the changes in surface composition and its link with the current surface composition of the Sun. Therefore, this relation intrinsically includes a model-dependent dimension through the hypotheses made on the physical ingredients of the solar models, the most straightforward one being mixing prescriptions for chemicals.  

Early works by \citet{Serenelli2010} estimated the protosolar helium abundance from the analysis of a large number of Standard Solar Models (SSMs). They also included in their work some early models simulating turbulence at the base of the solar convective envelope. This work however only considered one estimate of the surface helium abundance \citep{BasuYSun, BasuY2004}, that has since been revised using different equations of state \citep{Vorontsov13,VorontsovSolarEnv2014,Buldgen2024Z}, and did not include the depletion of lithium and beryllium when studying the evolution of the surface helium abundance. They however considered a $20\%$ uncertainty in microscopic diffusion and related the primordial helium abundance to the surface one using scaling laws from \citet{Bahcall1989}. The uncertainties on the treatment of microscopic diffusion have since improved, thanks to full treatment of the effects and extensive comparisons of solar models \citep{Turcotte,Deal2025}. We also refer to \cite{Michaud2015} for a more in-depth discussion on microscopic diffusion. 

In this work, we take advantage of the recent determination of lithium ($\rm{A(Li)}=0.96\pm0.05$) \citep{Wang2021} and beryllium ($\rm{A(Be)}=1.21\pm0.05$) \citep{Amarsi2024} which are key tracers of the efficiency of the turbulent mixing at the base of the convective zone (BCZ) as these light elements are depleted when compared to meteoritic values, $\rm{A(Li)} = 3.27\pm0.03$ and $\rm{A(Be)} = 1.31\pm0.04$ \citep{Lodders2021}. In previous works \citep{Buldgen2023,Buldgen2025Be}, we showed that including a calibrated mixing efficiency on both elements strongly impacted the initial helium abundance of solar models as well as the conclusions one could draw when comparing solar models to helioseismic constraints. In \citet{Deal2025}, we showed that these results were also obtained with various stellar evolution codes and calibration procedures. Using various ingredients that affect the helium mass fraction evolution  (opacities, overshooting, nuclear rates) within the same numerical framework, we aim to provide a robust estimation of the primordial helium abundance. 

We start in Sect. \ref{Sec:Models} by discussing the various physical hypotheses that may influence the primordial helium value, separating what concerns the transport of chemicals and other ingredients that will indirectly influence the evolution of the helium mass fraction in the envelope. From this analysis, we can draw an estimate of the difference between the primordial and current envelope helium mass fraction in Sect. \ref{Sec:PhyIng} and finally an estimate of the primordial helium mass fraction in Sect. \ref{Sec:EstimateHe}. We conclude by discussing additional improvements to solar models that may help us further increase the precision of this estimate in the coming years.

\section{Solar Models}\label{Sec:Models}

In this section, we present the properties of the solar models computed with the Liège Stellar Evolution Code \citep{ScuflaireCles} that we will use to infer the primordial solar helium abundance, $\rm{Y_{P}}$. We use the AAG21 solar abundances \citep{Asplund2021}, the OPAL \citep{OPAL} and LANL/OPLIB \citep{Colgan} opacities, the mixing length formalism of convection following \citet{CoxGiuli1968}, the SAHA-S equation of state (version 7) and the NACRE II nuclear reaction rates \citep{Xu2013}. Our models include the \citet{Thoul} formalism of diffusion, including the \citet{Paquette} screening coefficients and consider partial ionization of the chemical elements. Turbulence is treated following \citet{Proffitt1991} as in \citet{Buldgen2025Be}. However, given the similarities found in \citet{Buldgen2025Be} and \citet{Deal2025} for other analytical expressions of the empirical turbulent diffusion coefficient, this does not constitute a limitation on the conclusions we draw on turbulence at the BCZ. 

The main properties of our solar models are summarized in Tables \ref{tabModelsDTurb}, considering first only the effects of turbulent diffusion at the BCZ and \ref{tabModelsOpOvDTurb} where we consider a change of reference opacities, overshooting and turbulent diffusion. The results for the models considering both the effects of overshooting and turbulent diffusion over a large range of parameters are provided in Table  \ref{tabModelsOvDTurb}. We use the following naming convention: $D^{X}_{Y}$ implies the inclusion of an additional diffusion coefficient in $cm^{2}/s$ of the form, as in \citet{Proffitt1991}:
\begin{align}
D^{X}_{Y}= Y \left(\frac{\rho_{\rm{BCZ}}}{\rho}\right)^{X}, \label{eq:Proff}
\end{align}
with $\rho_{\rm{BCZ}}$ the density at the BCZ position, $\rho$ the local density. $X$ and $Y$ are the free parameters in the coefficient that have to be calibrated (usually on light element depletion). \\

The addition of the suffix ``LANL'' implies the use of the Los Alamos / OPLIB opacities \citep{Colgan}, while ``Ov $Z$'' implies the inclusion of overshooting at the BCZ, over a fraction of the local pressure scale height defined by $Z$, enforcing instantaneous mixing of chemicals and an adiabatic temperature gradient in the overshooting region.  

\begin{figure}
	\centering
		\includegraphics[width=7.2cm]{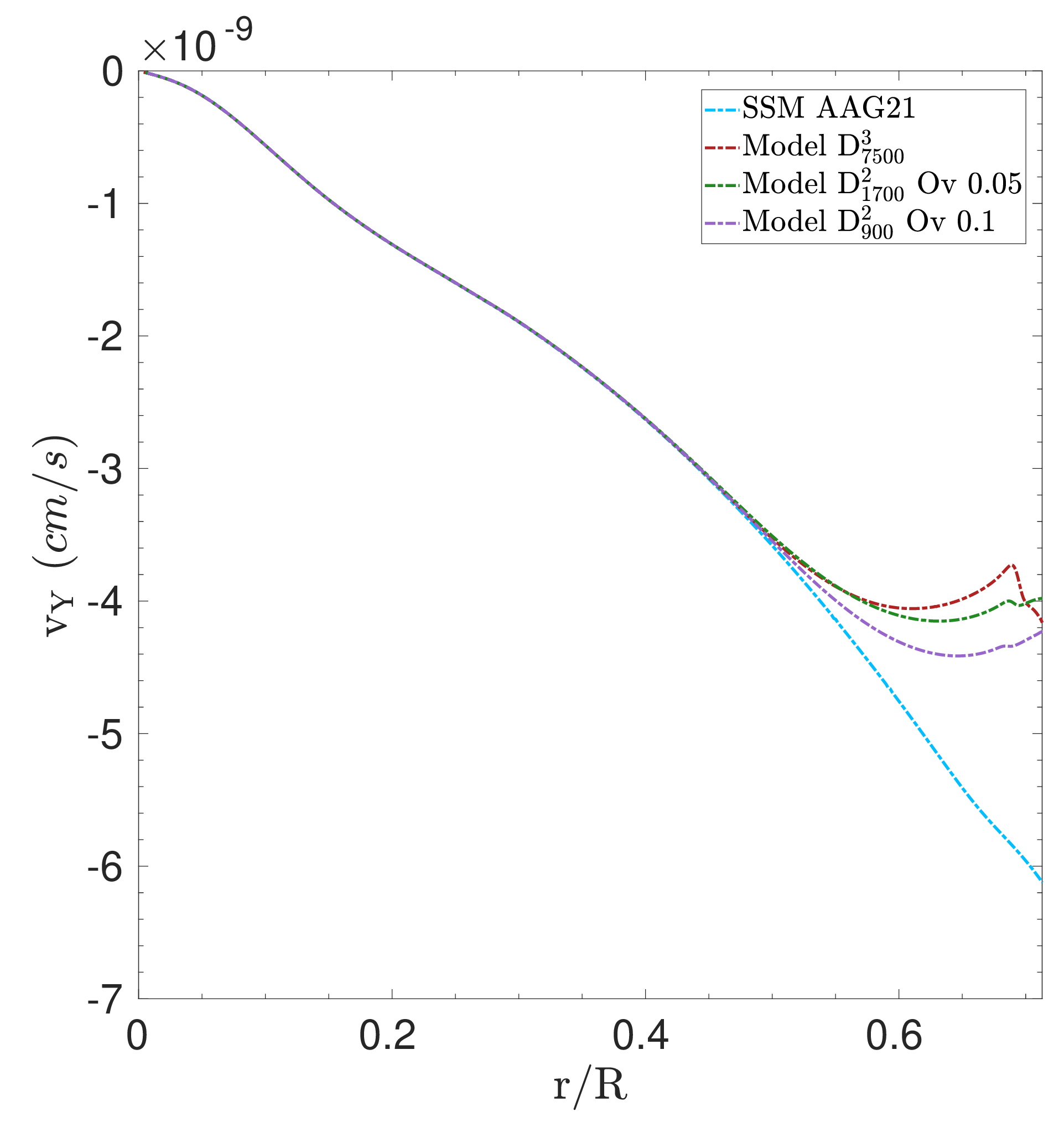}
	\caption{Diffusion velocity of helium for as a function of normalized radius in the radiative zone of solar models (a standard solar model in light blue and models including macroscopic mixing and/or overshooting at the BCZ in red, green and purple).}
		\label{Fig:VeloDiff}
\end{figure}

One can already see that directly inferring $\rm{Y_{P}}$ from calibrated solar evolutionary models is not feasible given the interdependencies between the physical ingredients of solar models \citep[see e.g.][and refs therein for a recent review]{JCD2021}. For example, changing the opacity tables in solar models will directly affect $\rm{Y_{P}}$ as a change of opacity in the core of the Sun needs to be compensated such that the solar calibrated model reproduces the solar luminosity at the solar age \citep[see e.g.][for a discussion for a wide range of ingredients of solar models]{Buldgen2019}. However, we can already see a crucial element of the evolution of the chemical composition of the solar envelope. Namely, looking at the slope of the values of $\rm{Y_{CZ}}$, two effects stand out. First, the intensity of turbulence at the BCZ strongly affects the efficiency of settling (For example in the SSM, the value of $\rm{Y_{CZ}}$ is significantly lower). Second, the physical conditions at the BCZ (e.g. its position, the temperature gradients, etc) will also slightly affect the helium depletion over time, although on a much smaller scale. Indeed, microscopic diffusion effects are a combination of temperature, pressure and chemical composition contributions \citep[e.g.][]{Turcotte,Baturin2006}. We will illustrate this by plotting in Fig. \ref{Fig:VeloDiff} the diffusion velocity of helium for three solar models of our sample. 

\begin{table*}[h]
\caption{Global parameters of the solar evolutionary models including turbulent diffusion.}
\label{tabModelsDTurb}
  \centering
\begin{tabular}{r | c | c | c | c | c }
\hline \hline
\textbf{Name}&\textbf{$\left(r/R\right)_{\rm{BCZ}}$}&\textbf{$\mathit{Y}_{\rm{CZ}}$}&\textbf{$\mathit{Y}_{\rm{P}}$}&\textbf{$\mathit{A(Li)}$ (dex)}&\textbf{$\mathit{A(Be)}$ (dex)} \\ \hline
SSM &$0.7263$&$0.2371$& $0.2653$ & $3.28$ &$1.36$\\
D$^{2}_{1500}$&$0.7250$&$0.2469$& $0.2650$ & $1.87$& $1.22$\\
D$^{2}_{1700}$&$0.7251$&$0.2471$& $0.2649$ & $1.78$& $1.20$\\
D$^{2}_{2000}$&$0.7252$&$0.2473$& $0.2648$ & $1.66$&$1.18$\\
D$^{2}_{2500}$&$0.7254$&$0.2477$& $0.2646$ & $1.46$&$1.13$\\ 
D$^{2}_{3500}$&$0.7256$&$0.2483$& $0.2643$ & $1.11$&$1.05$\\ 
D$^{2}_{4000}$&$0.7257$&$0.2486$& $0.2642$ & $0.94$&$1.01$\\ 
D$^{2}_{4500}$&$0.7258$&$0.2488$& $0.2642$ & $0.78$&$0.98$\\
D$^{2}_{5500}$&$0.7260$&$0.2491$& $0.2641$ & $0.48$&$0.91$\\
D$^{3}_{6500}$&$0.7255$&$0.2481$& $0.2644$ & $0.97$&$1.15$\\
D$^{3}_{6700}$&$0.7255$&$0.2481$& $0.2644$ & $0.94$&$1.14$\\
D$^{3}_{7500}$&$0.7256$&$0.2483$& $0.2643$ & $0.80$&$1.12$\\

\hline
\end{tabular}
\end{table*}

\begin{table*}[h]
\caption{Global parameters of the solar evolutionary models using the OPLIB opacities, including overshooting and turbulent diffusion.}
\label{tabModelsOpOvDTurb}
  \centering
\begin{tabular}{r | c | c | c | c | c }
\hline \hline
\textbf{Name}&\textbf{$\left(r/R\right)_{\rm{BCZ}}$}&\textbf{$\mathit{Y}_{\rm{CZ}}$}&\textbf{$\mathit{Y}_{\rm{P}}$}&\textbf{$\mathit{A(Li)}$ (dex)}&\textbf{$\mathit{A(Be)}$ (dex)} \\ \hline
LANL Ov $0.17$ D$^{1}_{750}$&$0.7122$&$0.2383$& $0.2557$ & $1.48$&$1.17$\\
LANL Ov $0.17$ D$^{1}_{850}$&$0.7123$&$0.2386$& $0.2556$ & $1.41$&$1.14$\\
LANL Ov $0.17$ D$^{2}_{1250}$&$0.7120$&$0.2383$& $0.2556$ & $1.38$&$1.22$\\
LANL Ov $0.1$ D$^{2}_{1150}$&$0.7176$&$0.2380$& $0.2558$ & $1.83$ & $1.24$\\
LANL Ov $0.15$ D$^{2}_{1450}$&$0.7137$&$0.2385$& $0.2555$ & $1.41$ & $1.20$\\
LANL Ov $0.12$ D$^{2}_{1750}$&$0.7163$&$0.2387$& $0.2554$ & $1.45$&$1.18$\\
LANL Ov $0.22$ D$^{2}_{1250}$&$0.7079$&$0.2385$& $0.2555$ & $1.01$&$1.21$\\
LANL Ov $0.25$ D$^{2}_{1150}$&$0.7053$&$0.2385$& $0.2555$ & $0.81$&$1.21$\\
\hline
\end{tabular}
\end{table*}

As can be seen, the presence of turbulence directly affects the diffusion velocity at the BCZ. The differences between the SSM and the model including turbulence from \citet{Proffitt1991} go as high as 30$\%$, explaining the observed differences in $\rm{Y_{CZ}}$ at the age of the Sun. This effect is visible even for lower intensity of turbulent diffusion, even when overshooting is considered and already well before the models reproduce the lithium and/or beryllium depletion. This implies that such a reduction of the efficiency of microscopic diffusion remains relevant even if the uncertainties of the current lithium and beryllium abundances were underestimated.

\begin{figure}
	\centering
		\includegraphics[width=7.8cm]{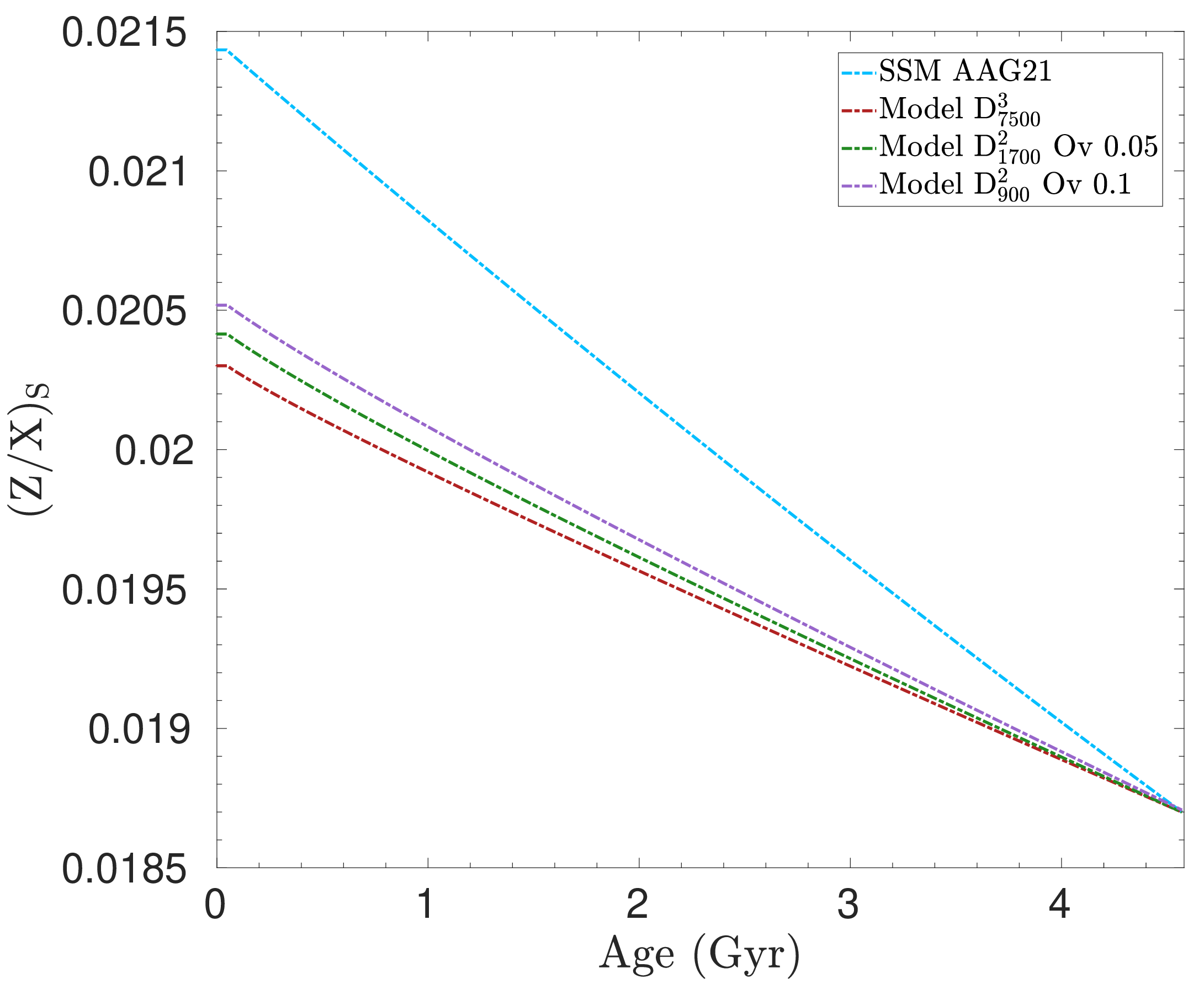}
	\caption{Evolution of the surface metallicity of calibrated solar models, $(Z/X)_{S}$, as a function of time (a standard solar model in light blue and models including macroscopic mixing and/or overshooting at the BCZ in red, green and purple). The assumed final metallicity at the solar age is that of \citet{Asplund2021}.}
		\label{Fig:ZXEvol}
\end{figure}

As it impacts the transport of all elements during the evolution of the Sun, turbulent diffusion at the BCZ has a direct impact on the solar calibration procedure. Indeed, a calibrated solar model must reproduce the solar radius, luminosity and surface metallicity at the age of the Sun. Due to the effect of turbulence, the evolution of the metals is also impacted, meaning that the initial conditions of the solar calibration are affected by the presence of extra-mixing or by variations of the BCZ position. We illustrate this well-known effect in Fig. \ref{Fig:ZXEvol} with the evolution of the surface metallicity, which has been reported in numerous publications \citep[see e.g.][]{Richard1996,Gabriel1997,Brun2002,JCD2018,Buldgen2025Be}. The fact that the BCZ position can be precisely located using helioseismology strongly constrains the allowed variations and therefore the impact on the chemical evolution of the position of the convective envelope. 

\subsection{Impact of light element depletion on helium mass fraction}\label{Sec:LiBeHe}

Reproducing the observed depletion of lithium and beryllium is a crucial component of the chemical evolution of the Sun and Sun-like stars. While the exact underlying physical mechanism is still unknown, as recent candidates \citep{Eggenberger2022} are unable to reproduce lithium and beryllium simultaneously \citep{Buldgen2025Be}, it is clear that the observed depletion is due to additional mixing occuring at the BCZ. Previous works \citep{Schlattl1999,Zhang2019} have shown that the depletion of lithium can be reproduced by overshooting, but this process does not induce any beryllium depletion \citep{Kunitomo2025}, being thus at odds with observations \citep{Amarsi2024}. Results in \citet{Eggenberger2022} have also shown that overshooting induced a too high depletion of lithium in young solar twins in open clusters, implying that its impact should be limited. Additionally, \citet{Buldgen2025S} showed that overshooting cannot reproduce simultaneously the properties of the CZ such as the BCZ position and the height of the entropy plateau in the CZ, while an opacity increase at the BCZ can. These results tend to favour recent hydrodynamical simulations advocating for a limited extent of overshooting at the BCZ. In this context, solar models must reproduce simultaneously the BCZ position, the entropy plateau height, the lithium and the beryllium depletions. Such models can only be produced by assuming ad-hoc modifications to the opacity profile and/or additional phenomena/modifications to the physical ingredients \citep[see e.g. amongst others][]{Guzik06,JCD2009,Serenelli2009,Ayukov2011,Ayukov2017,Buldgen2019,Zhang2019,Kunitomo2021,Kunitomo2022}. 

\begin{figure}
	\centering
		\includegraphics[width=7.2cm]{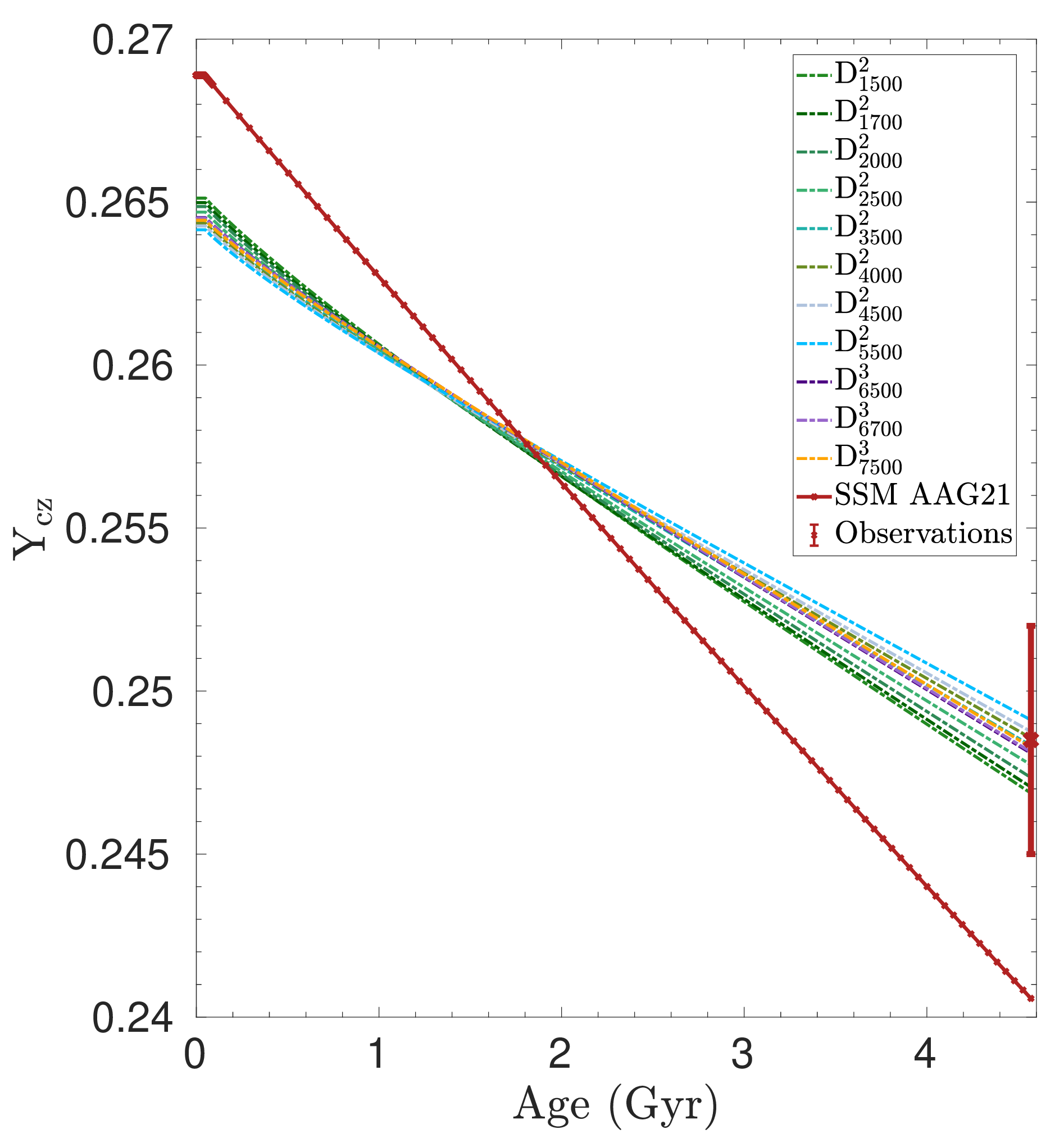}
	\caption{Evolution of the helium mass fraction in the convective envelope as a function of time for the models of Table \ref{tabModelsDTurb}. The ``Observed'' value (red cross) is taken as that of \citet{BasuY2004}.}
		\label{Fig:YCZDT}
\end{figure}

Even considering  the lithium and beryllium depletions as relatively loose constraints, the overall variation of $\rm{Y_{CZ}}$, presented in Fig. \ref{Fig:YCZDT}, remains quite similar for a fixed set of physical ingredients. The difference from the one of an SSM (shown here in red and denoted SSM AAG21) are striking, with a lower $\rm{Y_{P}}$ found in models including turbulence. 

\begin{figure*}
	\centering
		\includegraphics[width=14.6cm]{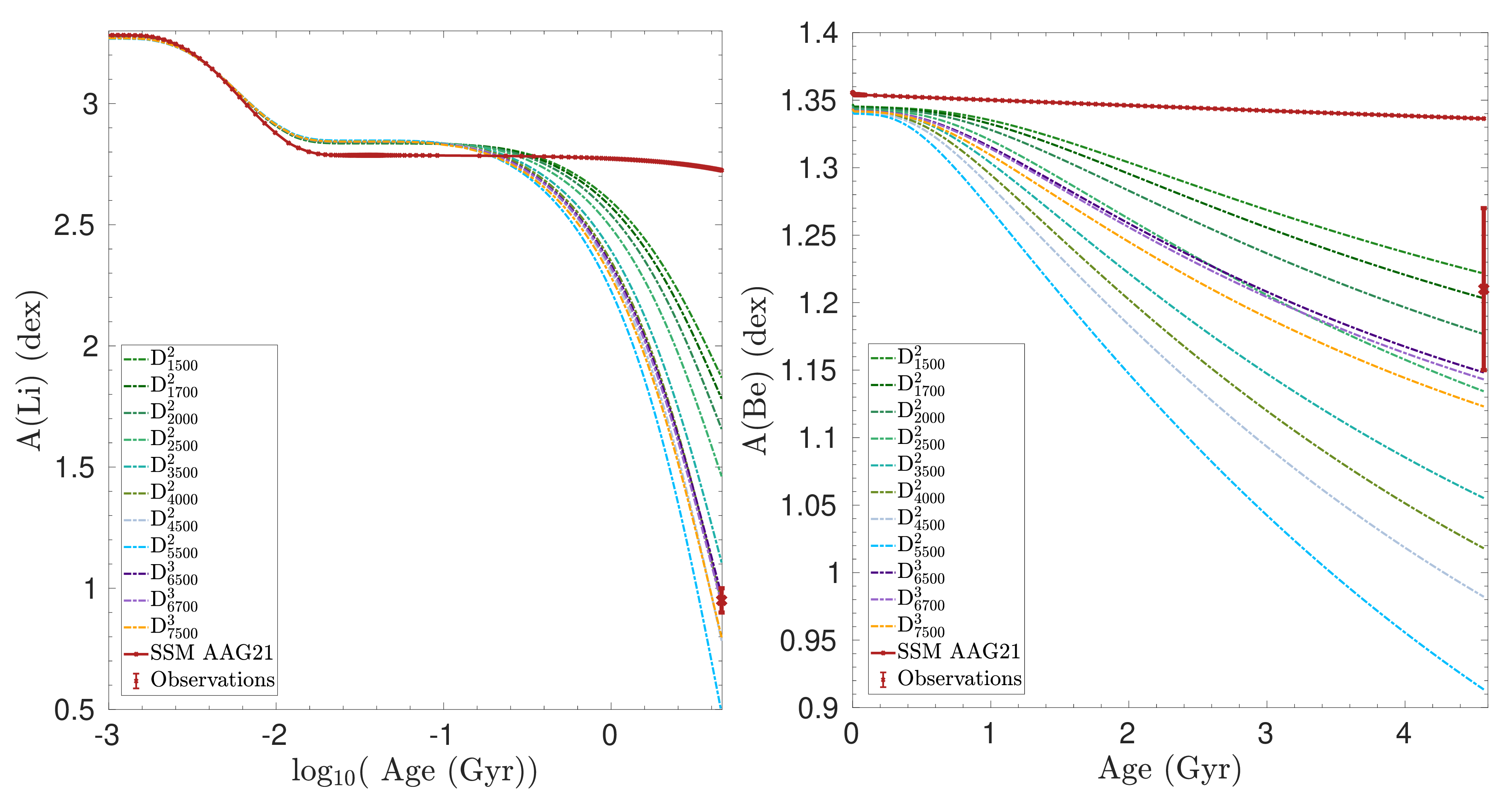}
	\caption{Left panel: Evolution of surface Lithium abundance as a function of age (in log scale) for the models of Table \ref{tabModelsDTurb}. The ``Observed'' value is taken from \citet{Wang2021}. Right panel: Evolution of the surface Beryllium abundance as a function of age (in log scale) for the models of Table \ref{tabModelsDTurb}. The ``Observed'' value is taken from \citet{Amarsi2024}.}
		\label{Fig:LiBeDT}
\end{figure*}

We illustrate in Fig. \ref{Fig:LiBeDT} the corresponding depletion of lithium and beryllium as a function of time in these models. As mentioned above, we did not consider the lithium and beryllium depletion to be strong constraints and allowed for models in stark disagreement with observations. Nevertheless, even by considering models with significantly lower depletion of light elements then the ones observed, the effect of turbulence of $\rm{Y_{CZ}}$ and $\rm{Y_{P}}$ was immediate. This is a direct consequence of the impact of turbulence on the diffusion velocities, meaning that the snapshot illustrated at the solar age in Fig. \ref{Fig:VeloDiff} where the efficient settling is damped already happens at much lower efficiencies of turbulent mixing at the BCZ and throughout the evolution of solar models. This demonstrates that the protosolar helium abundance cannot be inferred without considering the impact of turbulence at the BCZ, at it would induce a significant overestimation of its value.  

\subsection{Importance of physical ingredients of solar models}\label{Sec:PhyIng}

\begin{figure*}
	\centering
		\includegraphics[width=7.45cm]{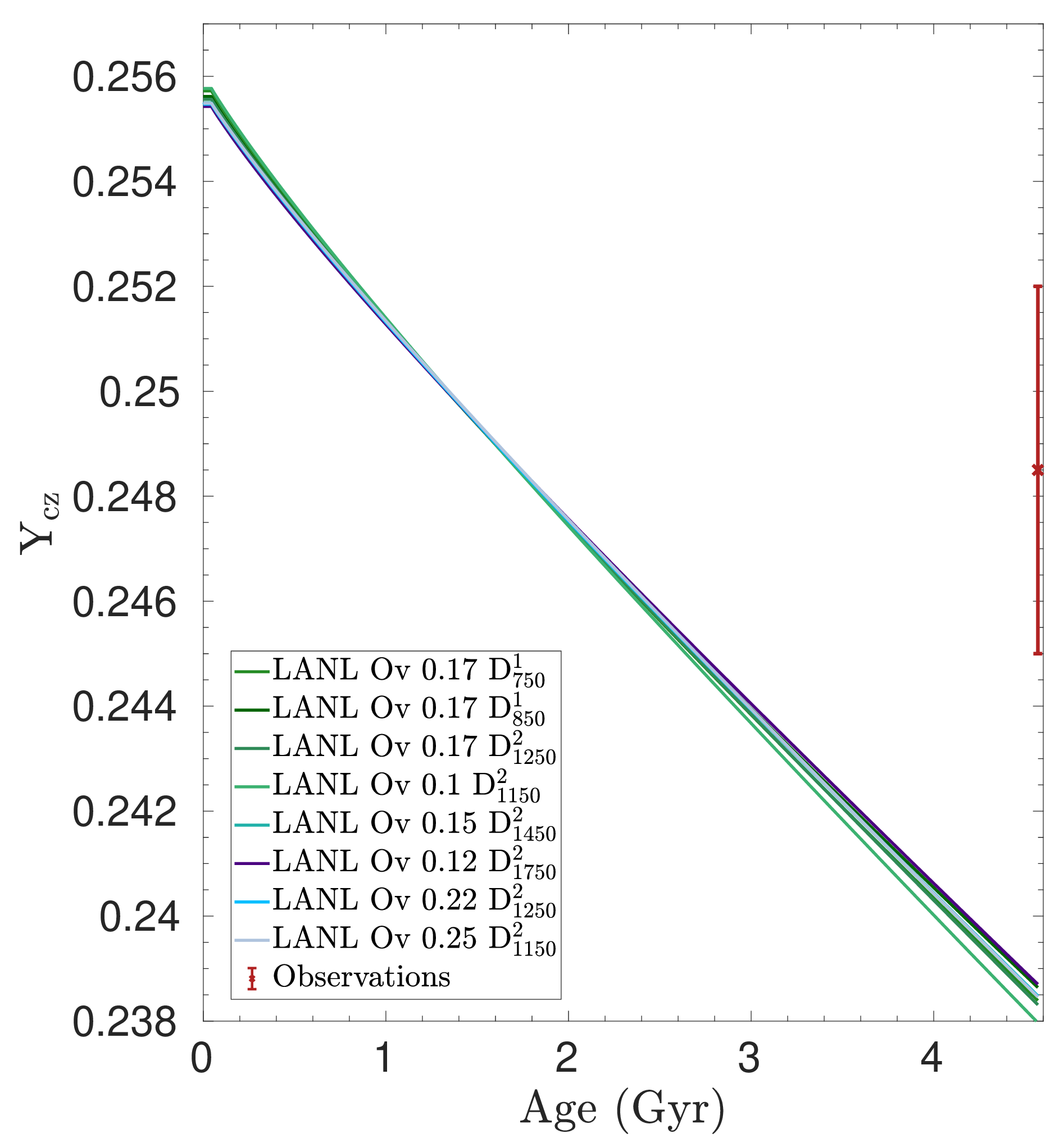}
		\includegraphics[width=7.34cm]{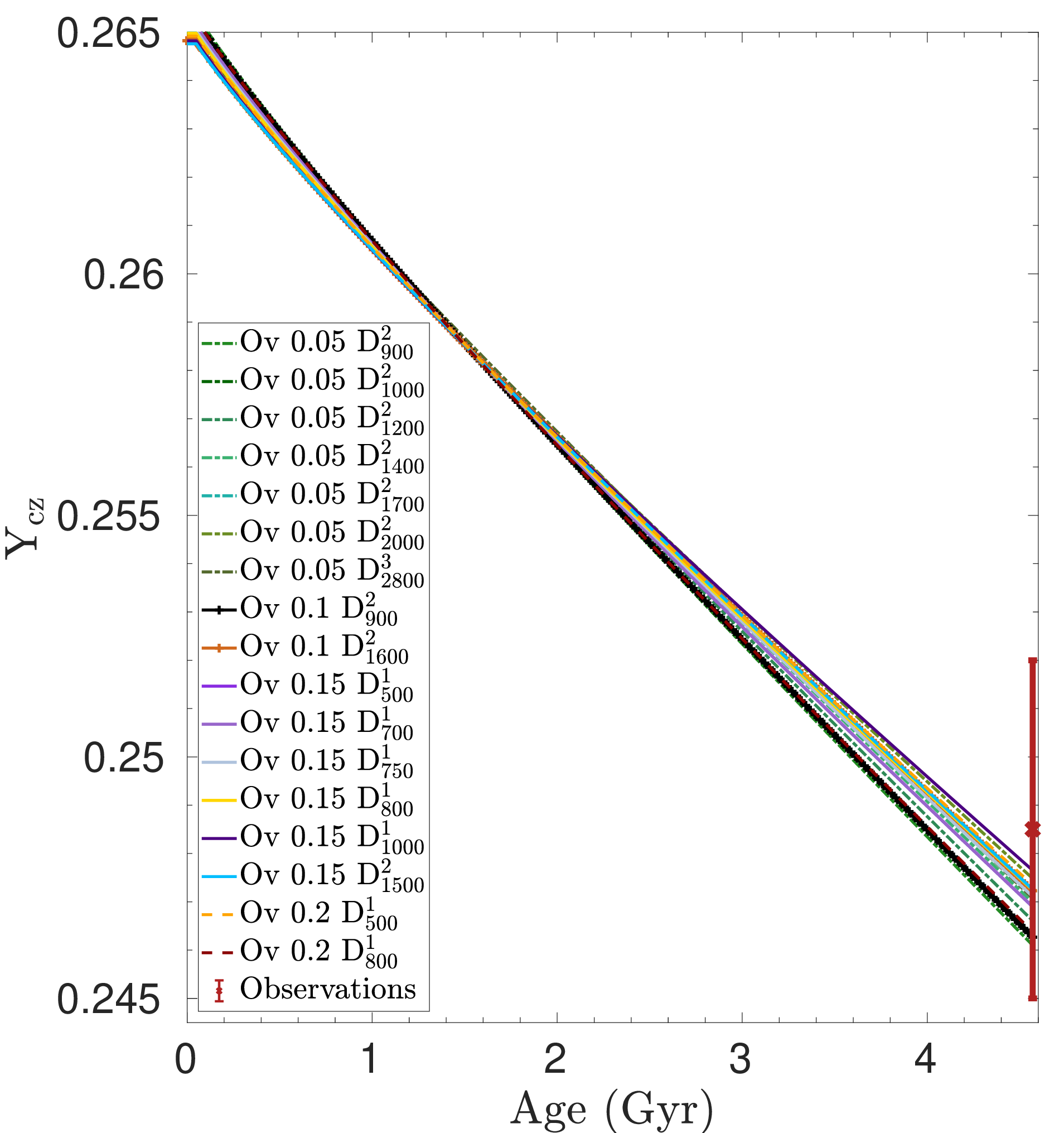}
	\caption{Left panel: Evolution of the helium mass fraction in the convective envelope as a function of time for the models of Table \ref{tabModelsOpOvDTurb}. The ``Observed'' value (red cross) is taken as that of \citet{BasuY2004}. Right panel:Evolution of the helium mass fraction in the convective envelope as a function of time for the models of Table \ref{tabModelsOvDTurb}.}
		\label{Fig:YLanlOV}
\end{figure*}

As mentioned before, physical ingredients of solar models may play a crucial role. Radiative opacities at high temperatures have a significant impact on $\rm{Y_{P}}$ with the latest OPLIB opacities leading to a lower value. This change is a direct consequence of a variation of the central temperature of the Sun induced by the intrinsically lower OPLIB opacities in this regime. A similar effect was observed in \citet{Ayukov2017} when changing the efficiency of nuclear reactions, or the opacity at high temperatures \citep[see][and refs therein for a detailed discussion]{JCD2021}. This effect is illustrated in the left panel of Fig. \ref{Fig:YLanlOV}, where we can see the evolution of $\rm{Y_{CZ}}$.

In addition to the effect of a change in opacity, we considered the impact of overshooting at the BCZ. As shown in various publications, adding overshooting leads to an enhanced depletion of lithium, particularly during the PMS phase. While this is at odds with young solar twins in open clusters, we still consider the combined impact of overshooting and turbulent diffusion at the BCZ to investigate its potential impact on the evolution of $\rm{Y_{CZ}}$. If we compare the right panel of Fig. \ref{Fig:YLanlOV} to Fig. \ref{Fig:YCZDT}, we can see that the impact is negligible. In other words the evolution of the BCZ position under the effect of overshooting has no influence on the depletion of helium, it is solely dominated by turbulence at the BCZ.  

We also illustrate in Figs. \ref{Fig:LiBeOPDT} and \ref{Fig:LiBeOVDT} the evolution of the surface lithium and beryllium abundances as a function of time for the models of Tables \ref{tabModelsOvDTurb} and \ref{tabModelsOpOvDTurb}. While the models cover a wide range of final values for both lithium and beryllium, their evolution of $\rm{Y_{CZ}}$ remains similar, implying that the dominant factor is purely the presence of turbulent diffusion at the BCZ. This remains true for a wide range of overshooting values and under the effect of a change of reference opacity tables. Indeed, the only effect of changing the OPAL opacity tables with the OPLIB opacity tables is a vertical shift of $\rm{Y_{P}}$ that is a simple consequence of the initial conditions of the calibration procedure that requires the calibrated model to reproduce the solar luminosity at the solar age (amongst the other constraints). A direct consequence of the left panel of Fig. \ref{Fig:YLanlOV} is that non-standard models reproducing both lithium and beryllium depletion using the OPLIB opacity tables will not be able to reproduce the $\rm{Y_{CZ}}$ values inferred from helioseismology when using the AAG21 abundances\footnote{This issue had already been discussed extensively in \citet{Buldgen2019} for a wide variety of physical ingredients.}. The problem is less stringent if a higher solar metallicity is considered \citep[e.g. using the MB22 abundances from][]{Magg2022}, but as shown in \citet{Buldgen2023} and \citet{Buldgen2024}, a strong disagreement remains with neutrino fluxes, confirming that uncertainties on radiative opacities might not only be an issue at temperatures of the BCZ.

\section{Evolution of the helium mass fraction in the convective envelope}\label{Sec:DYEvol}

As shown in Sect. \ref{Sec:PhyIng}, the absolute value of helium abundance in the CZ may vary depending on numerous physical ingredients, but the BCZ position itself has little impact. As already demonstrated in previous works \citep{Serenelli2010}, the only factors determining the value of $\rm{Y_{CZ}}$ are the initial conditions (thus $\rm{Y_{P}}$) and the efficiency with which helium settles from the BCZ. In this respect, we emphasized in the previous sections the crucial role of turbulent diffusion at the BCZ. We highlighted that it would significantly affect the settling of helium even without considering the current lithium and beryllium abundances as strong constraints in the calibration. 

\begin{figure*}
	\centering
		\includegraphics[width=7.4cm]{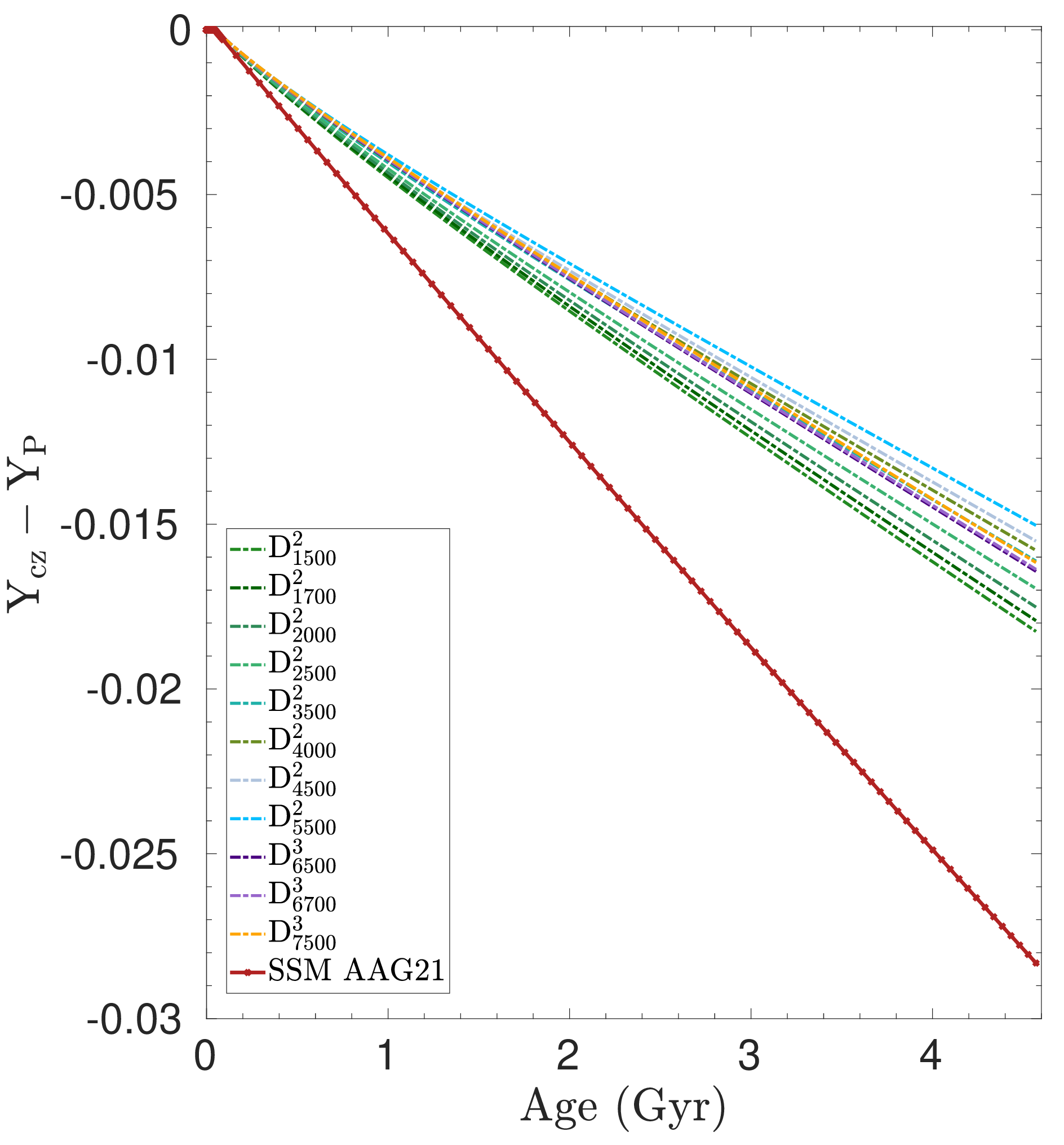}
		\includegraphics[width=7.4cm]{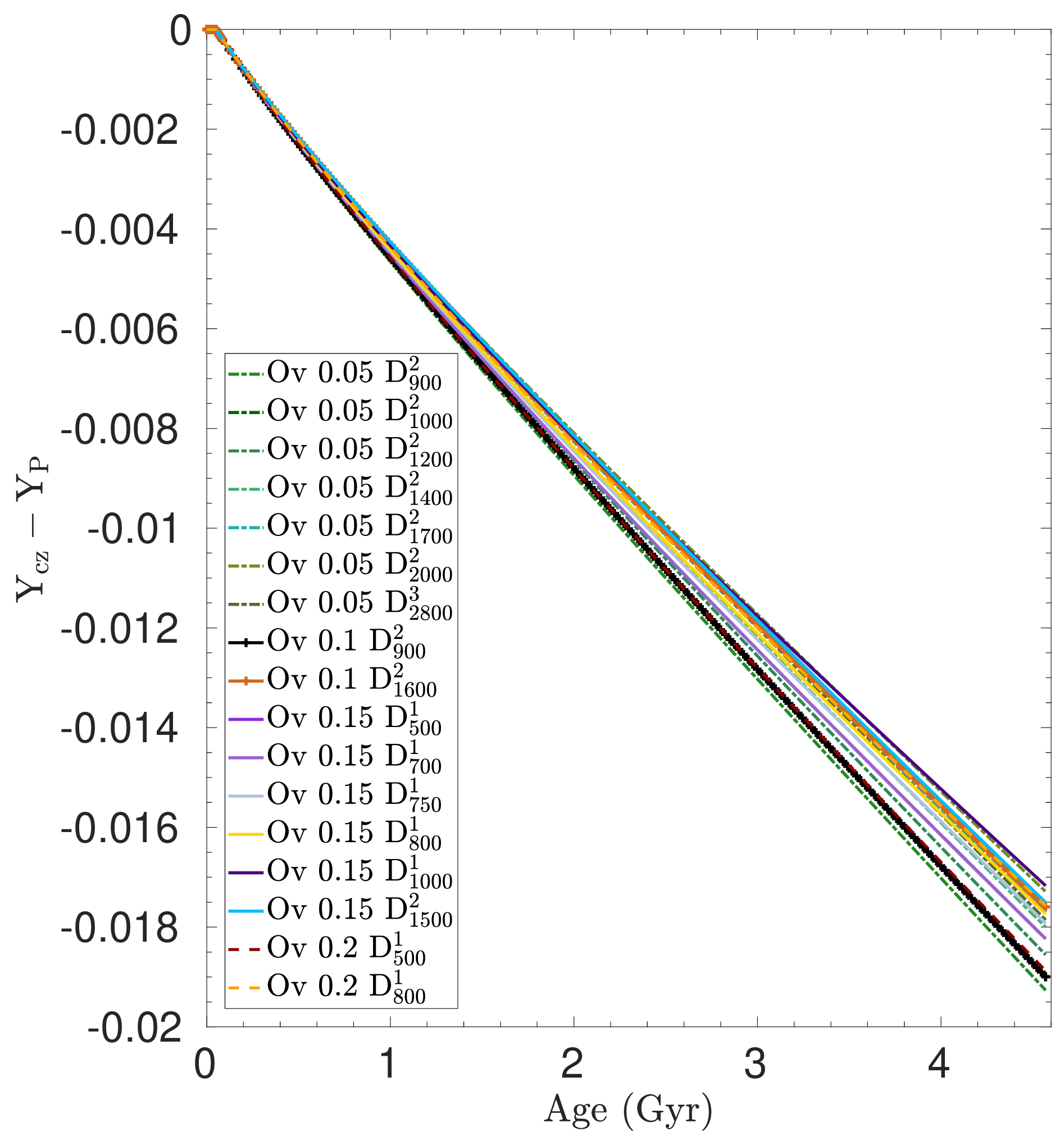}
	\caption{Left panel: Evolution of the difference between the protosolar and surface abundance of helium ($\rm{Y_{CZ}}$-$\rm{Y_{P}}$) as a function of time for the models of Table \ref{tabModelsDTurb}. Right panel: Same as the left panel for the models of Table \ref{tabModelsOpOvDTurb}.}
		\label{Fig:DiffYDTOv}
\end{figure*}

\begin{figure}
	\centering
		\includegraphics[width=7.4cm]{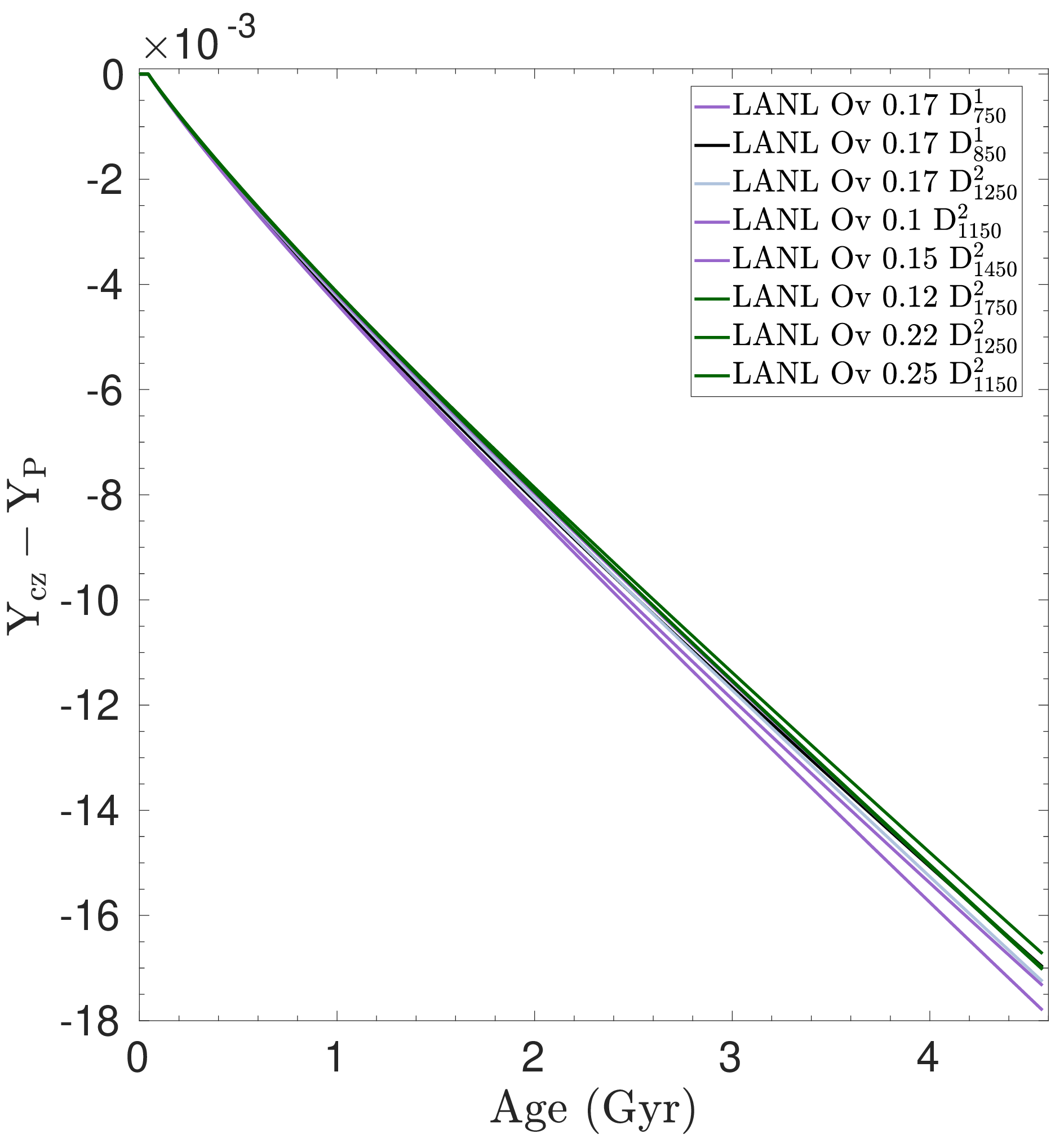}
	\caption{Evolution of the difference between the protosolar and surface abundance of helium ($\rm{Y_{CZ}}$-$\rm{Y_{P}}$) as a function of time for the models of Table \ref{tabModelsOpOvDTurb}.}
		\label{Fig:DiffYDTOp}
\end{figure}

To further illustrate this point, we plot in Figs. \ref{Fig:DiffYDTOv} and \ref{Fig:DiffYDTOp} the differences between the initial and primordial helium abundance $\rm{Y_{CZ}}$-$\rm{Y_{P}}$ as a function of time. In all cases the variation is quite similar and much smaller than the one obtained with a SSM, illustrated in the left panel of Fig. \ref{Fig:DiffYDTOv}. The same trend is always obtained, whatever the physical ingredients, the more efficient the turbulent diffusion at the BCZ, the lower the efficiency of settling and the lower the lithium value and beryllium at the age of the Sun. An important point is that the effect of the opacities has been completely erased, as the observed differences, $\rm{Y_{CZ}}$-$\rm{Y_{P}}$, for models using OPLIB opacities or OPAL opacities remain well within the same range. This confirms that we can directly infer $\rm{Y_{P}}$ from the knowledge of $\rm{Y_{CZ}}$ and the efficiency of the transport at the BCZ quantified through $\rm{Y_{CZ}}$-$\rm{Y_{P}}$ at the solar age. 

From a quantitative point of view, the value $\rm{Y_{CZ}}$-$\rm{Y_{P}}$ for the SSM is $-0.02831$, very similar to what was found in \citet{Serenelli2010}. As soon as turbulent diffusion is included, considering our entire set of models, this value drops within an interval between $-0.01523$ and $-0.1921$. If we consider the lithium and beryllium depletion as strong constraints, this interval further reduces to values between $-0.01750$ and $-0.01888$. This means that using a SSM instead of models taking into account light element depletion induces an overestimation of  $\rm{Y_{CZ}}$-$\rm{Y_{P}}$ by about $40\%$, significantly biasing the estimated protosolar helium abundance. 

\section{Estimated protosolar helium using helioseismology of the solar envelope}\label{Sec:EstimateHe}

\begin{figure*}
	\centering
		\includegraphics[trim= 0 0 0 0, clip, width=14.5cm]{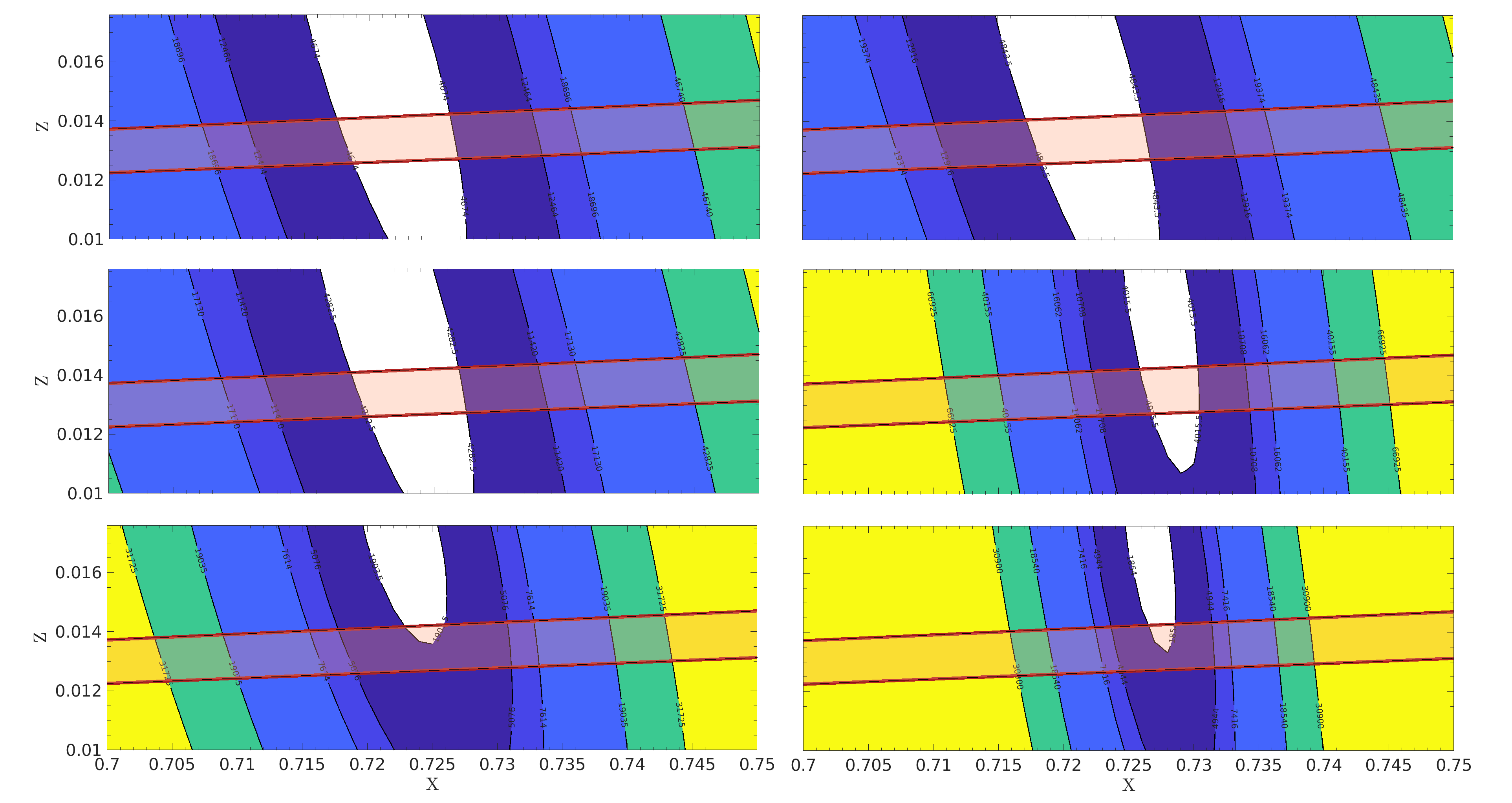}
	\caption{$\chi^{2}$ maps used in \citet{Buldgen2024Z} to determine the value of the hydrogen mass fraction as a function of the metal mass fraction in the solar convective envelope. The properties of the models are summarized in their table 3. The red band shows the range of values compatible with the AAG21  $(Z/X)_{S}$ value.}
		\label{Fig:MapProto}
\end{figure*}

In this Section, we will combine spectroscopic constraints, helioseismic constraints and the inferred properties of $\rm{Y_{CZ}}$-$\rm{Y_{P}}$ to recover an updated value of $\rm{Y_{P}}$. To do this, we make use of the $\Gamma_{1}$ inversions of \citep{Buldgen2024Z}, more specifically the inferences in the higher part of the convective envelope, from which the helium abundance can be estimated. We show in Fig. \ref{Fig:MapProto} an example of a $\chi^{2}$ map used in this paper over which we overlay the surface area covered when assuming a value of $Z/X=0.0187 \pm 0.0009$ in agreement with AAG21. In practice, and as shown in \citet{Serenelli2010}, the assumed $Z/X$ has little impact on the inferred $\rm{Y_{CZ}}$ using helioseismic inversions. It may be neglected to first order, but one should keep in mind that some signatures of the partial ionization of metals in the $\Gamma_{1}$ profile are also present at the temperatures were we infer the $\rm{Y_{CZ}}$ value \citep[see e.g.][for an illustration of the contributions of metals to the $\Gamma_{1}$ profile]{Baturin2022}, inducing a small interdependency between the inferred $\rm{Y_{CZ}}$ and $\rm{Z_{CZ}}$ (the convective zone metal mass fraction) that can be seen in \citet{Vorontsov13} and \citet{VorontsovSolarEnv2014}.

In order to perform some rudimentary global analysis, we summarize in Table \ref{tabEOSHelium} some $\rm{Y_{CZ}}$ inference results from the literature. These values span over three decades of observations, different instruments, datasets, methods and equations of state. Overall the range found is not too different between the various works. However, we can note that the equation of state largely dominates the  uncertainty budget in the inference. This underlines the much needed improvements of the solar equation of state as mentioned in \citep{Baturin2025,Trampedach2025}.
\begin{table*}[h]
\caption{Helium mass fraction inferred from various groups in the literature.}
\label{tabEOSHelium}
  \centering
\begin{tabular}{r | c | c }
\hline \hline
\textbf{Reference}&\textbf{$\rm{Y_{CZ}}$}&\textbf{Equation of state} \\ \hline
\citet{RichardY}&$0.2480 \pm 0.0020 $& OPAL\\
\citet{DiMauro2002}&$0.2457 \pm 0.0005 $& MHD\\
\citet{DiMauro2002}&$0.2539 \pm 0.0005 $& OPAL\\
\citet{BasuY2004}&$0.24875 \pm 0.0035 $& OPAL\\ 
\citet{Vorontsov13}&$0.2475 \pm 0.0075 $& OPAL and SAHA-S\\
\citet{VorontsovSolarEnv2014}&$0.2525 \pm 0.0075 $& OPAL and SAHA-S\\
This work &$0.2575 \pm 0.0025 $& FreeEOS and SAHA-S\\
\hline
\end{tabular}
\end{table*}

From our previous work \citep{Buldgen2024Z} combined with \citet{Asplund2021}, we find the following interval for $\rm{Y_{CZ}}$ from our $\chi^{2}$ maps, $0.255-0.26$, in line with the higher values of the results of \citet{Vorontsov13} and \citet{VorontsovSolarEnv2014}. The main contributor to the uncertainty is again the equation of state, as we considered both SAHA-S (v3 and v7) and FreeEOS \citep{Irwin} for these inferences. It appears that this interval encompasses most of the ones found in table \ref{tabEOSHelium}. If we now combine these values to the $\rm{Y_{CZ}}$-$\rm{Y_{P}}$ differences found in previous sections, we obtain the following result for each $\rm{Y_{CZ}}$, assuming strong constraints on Lithium and Beryllium depletion:
\begin{itemize}
\item For \citet{RichardY}: $\rm{Y_{P}}=0.26619\pm 0.00269$;
\item For \citet{DiMauro2002} (MHD): $\rm{Y_{P}}=0.26385\pm 0.00115$;
\item For \citet{DiMauro2002} (OPAL): $\rm{Y_{P}}=0.27205\pm 0.00115$;
\item For \citet{BasuY2004}: $\rm{Y_{P}}=0.2669\pm 0.00415$;
\item For \citet{Vorontsov13}: $\rm{Y_{P}}=0.26565\pm 0.00815$;
\item For \citet{VorontsovSolarEnv2014}: $\rm{Y_{P}}=0.27065\pm 0.00815$;
\item For this work : $\rm{Y_{P}}=0.27565\pm 0.00315$.
\end{itemize}

A very conservative estimate for $\rm{Y_{P}}$ would fall between $0.2552$ and $0.2792$, considering the lowest (highest) estimates of $\rm{Y_{CZ}}$, namely $0.24$ and $0.26$, combined with the lowest (highest) estimate of $\rm{Y_{CZ}}$-$\rm{Y_{P}}$, namely $-0.01523$ and $-0.1921$. If we consider only the most recent works and considering the depletion of lithium and beryllium as strong constraints to be achieved by solar models, this interval reduces to $\left[0.2725,0.2789\right]$. Our value is slightly lower, yet consistent with the interval found by \citet{Serenelli2010} $\left[0.272,0.284 \right]$ but for entirely different reasons. In our case, the $\rm{Y_{CZ}}$ is intrinsically higher than theirs due to the $\rm{Y_{CZ}}$ intervals found in the latest analyses, whereas our depletion of helium over time is lower as a result of the inhibition of settling by turbulence at the BCZ. 

In fact, had we considered their $\rm{Y_{CZ}}$ and our conservative $\rm{Y_{CZ}}$-$\rm{Y_{P}}$ values, we would have found $\rm{Y_{P}}$ to be within $\left[0.2605,0.271 \right]$. This is even lower than their value estimated from non-SSMs which in their case did not include light element depletion. In all cases, the uncertainties on the $\rm{Y_{CZ}}$ value largely dominate the total budget, with uncertainties on the equation of state dominating the uncertainties on $\rm{Y_{CZ}}$ itself \citep[see e.g.][for an early discussion]{Kosovichev1992}. 

It should also be underlined that the various references provided in Table \ref{tabEOSHelium} used different techniques to infer $\rm{Y_{CZ}}$, as well as different datasets from different instruments. This may give rise to small systematic differences as noted in \citet{BasuY2004} or \citet{Buldgen2024Z} who used revised MDI data. Another tedious point is that of surface effect corrections. Apart from the works of \citet{Vorontsov13} and \citet{VorontsovSolarEnv2014}, all other methods used directly the individual frequencies and thus had to rely on empirical corrections for the surface effects, following for example \citet{RabelloParam} or using a newly derived expression for high degree modes \citep{DiMauro2002}. It should also be noted that the differences between \citet{Vorontsov13} and \citet{VorontsovSolarEnv2014} stem from different techniques, as the former used two approaches based on phase-shift of frequencies while the latter used group velocities of the solar acoustic modes. In this context, the uncertainties provided on $\rm{Y_{CZ}}$ are directly related to the ones of the inversion technique and thus in the case of some technique mentioned above (as the SOLA technique in \citet{Buldgen2024Z}) only consider the propagation of the observational error bars on the frequencies and not other sources of errors.

Therefore, an improvement of $\rm{Y_{P}}$ would also require to improve the inference of $\rm{Y_{CZ}}$ and carry out a detailed meta-analysis of the overall procedure. In order words, to analyse in a consistent way the impact of various helioseismic datasets, various inference techniques and equations of state used to determine the value of $\rm{Y_{CZ}}$, as well as various transport prescriptions (using for example various stellar evolution codes) on the inferred value of $\rm{Y_{P}}$. Such an analysis is beyond the scope of this paper and would only be conclusive if the various equations of state available to solar modellers were to be tested within this framework, using MCMC or bootstrap techniques to provide robust uncertainties on the inferred $\rm{Y_{CZ}}$ value. 

\section{Conclusion}\label{Sec:Conc}

In this study, we have analyzed in details the dependencies of the inference of the protosolar helium mass fraction, $\rm{Y_{P}}$, aiming at determining an updated value. We have shown that using Standard Solar Models leads to a significant overestimation by about $40\%$ of the depletion of helium from the CZ during the solar evolution when compared to solar models reproducing light element depletion. This result is independent of the opacity tables used or of the inclusion of overshooting at the base of the convective zone. We have shown that the lithium and beryllium abundances at the solar age could serve as efficient calibrators of the efficiency of mixing at the BCZ during the main-sequence evolution of the Sun. Using these constraints, we have carried out a meta-analysis of all the values in the litterature and inferred an updated interval for $\rm{Y_{P}}$. A very conservative estimate would be $\rm{Y_{P}}=0.2672\pm0.012$. 

When only considering the values for $\rm{Y_{CZ}}$ from \citet{Buldgen2024Z} and lithium and beryllium as strong constraints, we find $\rm{Y_{P}}=0.2757\pm0.0032$. This demonstrates the importance of considering the constraints on chemical mixing at the BCZ, as it significantly increases the precision of the inferred $\rm{Y_{P}}$ value.

This value remains consistent with that of \citet{Serenelli2010} of $\rm{Y_{P}}=0.278\pm0.006$, but this is due to a compensation effect as the $\rm{Y_{CZ}}$ values inferred in recent works is higher than the one from \citep{BasuYSun} used in \citet{Serenelli2010}. Taking their value of $\rm{Y_{CZ}}$ and remaining conservative on the effects of lithium and beryllium depletion, we wound find $\rm{Y_{P}}=0.2669\pm 0.00415$. This difference underlines the importance of taking into account macroscopic mixing at BCZ, as one cannot assume that Standard Solar Models provide an accurate representation of the chemical evolution of the solar convective envelope as they only include microscopic diffusion. The quoted value of $\rm{Y_{P}}=0.2669\pm 0.00415$ is perfectly consistent with the value reported in the work of \citet{Kunitomo2025} that used an extended calibration scheme including protosolar accretion and the $\rm{Y_{CZ}}$ value of \citet{BasuY2004} and the OPAL equation of state to describe the solar plasma. It is also noteworthy that protosolar accretion does not affect the inferred value of $\rm{Y_{P}}$ unless the helium mass fraction of the accreted material is significantly different from the protosolar one \citep[such an hypothesis was made in][]{Zhang2019}.

Improving the precision of the inference of $\rm{Y_{P}}$ using solar models requires first to improve the uncertainties on the $\rm{Y_{CZ}}$ value itself. This requires further improvement of the accuracy of the equation of state of the solar plasma, as it is the main source of uncertainty in the helioseismic inference of $\rm{Y_{CZ}}$. The second factor influencing the $\rm{Y_{CZ}}$ is the depiction of the outer convective envelope and the so-called surface effect. Limiting their impact by using averaged 3D hydrodynamical simulations instead of grey atmosphere models would further improve the issue. The third limiting factor in the precision of $\rm{Y_{P}}$ is the transport of chemicals at the base of the convective zone. While the physical nature of the macroscopic mixing at the BCZ is unknown, its overall efficiency and impact can still be empirically calibrated using the observed depletion of lithium and beryllium. In this respect, an improved precision on the inference of the current beryllium abundance in the solar photosphere would directly impact our uncertainty on $\rm{Y_{P}}$. Indeed, the beryllium depletion is an important anchoring point for the efficiency of mixing during the main-sequence while lithium is partially burned during the pre-main sequence. 

The above statement does not imply that improvements of the way the transport of chemicals at the BCZ itself is modelled would not significantly increase the precision of the inferred value of $\rm{Y_{P}}$. A better understanding of the transport of chemicals in the Sun is in any case paramount to improve solar models in general and our inference of stellar ages in light of the requirements of the PLATO mission \citep{Rauer2025}. However, it is likely that the lithium and beryllium depletion will remain crucial observational constraints to confront models with improved transport of chemicals. 

\begin{acknowledgements}

We thank the referee for their careful reading of the manuscript. GB acknowledges fundings from the Fonds National de la Recherche Scientifique (FNRS) as a postdoctoral researcher. 

\end{acknowledgements}

\bibliography{biblioarticleProto}

 \begin{appendix}

\section{Additional table and figures}
\label{app:AddFig}

\begin{figure*}
	\centering
		\includegraphics[width=15cm]{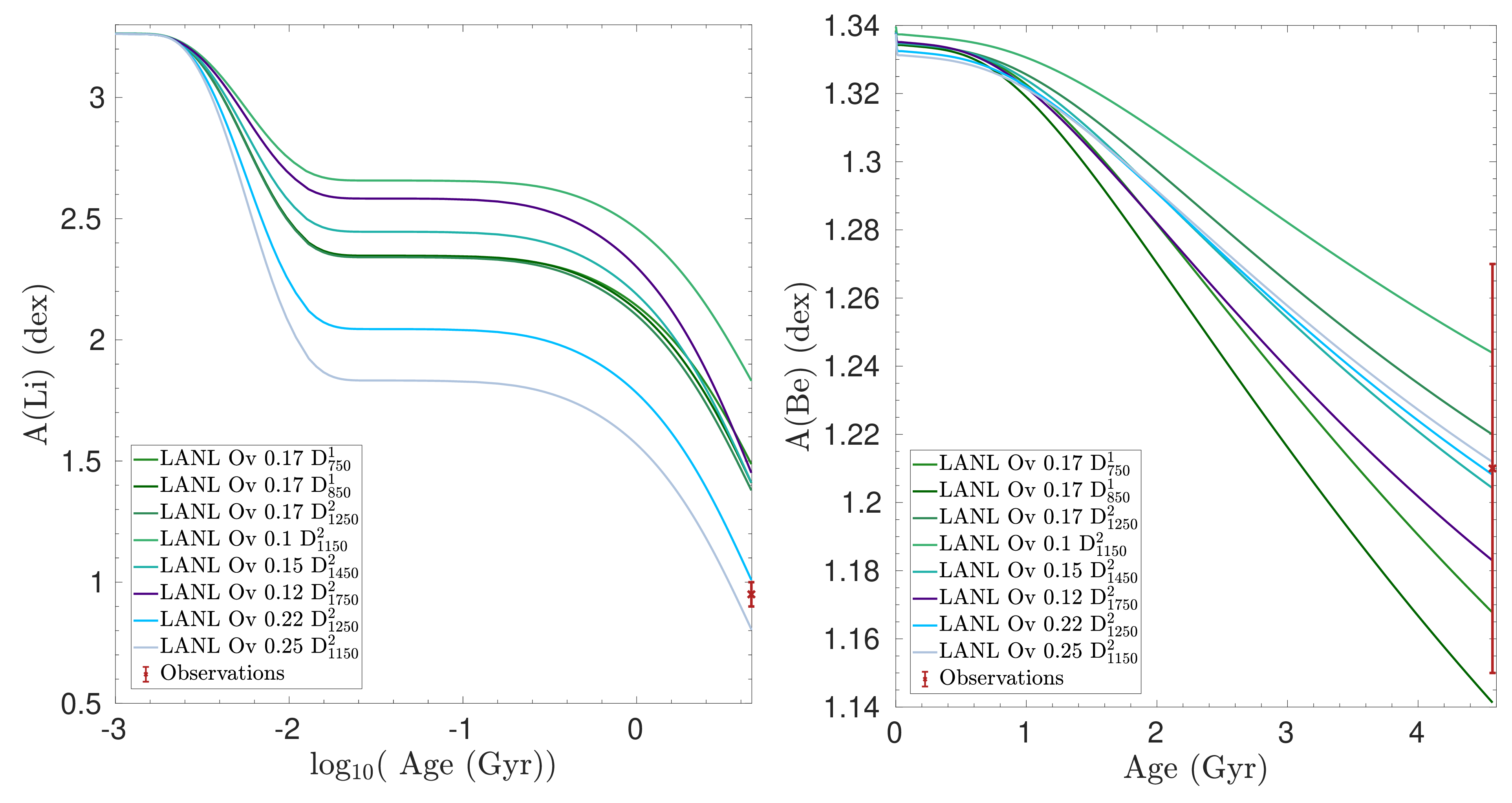}
	\caption{Left panel: Evolution of surface Lithium abundance as a function of age (in log scale) for the models of Table \ref{tabModelsOpOvDTurb}. The ``Observed'' value is taken from \citet{Wang2021}. Right panel: Evolution of the surface Beryllium abundance as a function of age (in log scale) for the models of Table \ref{tabModelsOpOvDTurb}. The ``Observed'' value is taken from \citet{Amarsi2024}.}
		\label{Fig:LiBeOPDT}
\end{figure*}

\begin{figure*}
	\centering
		\includegraphics[width=15cm]{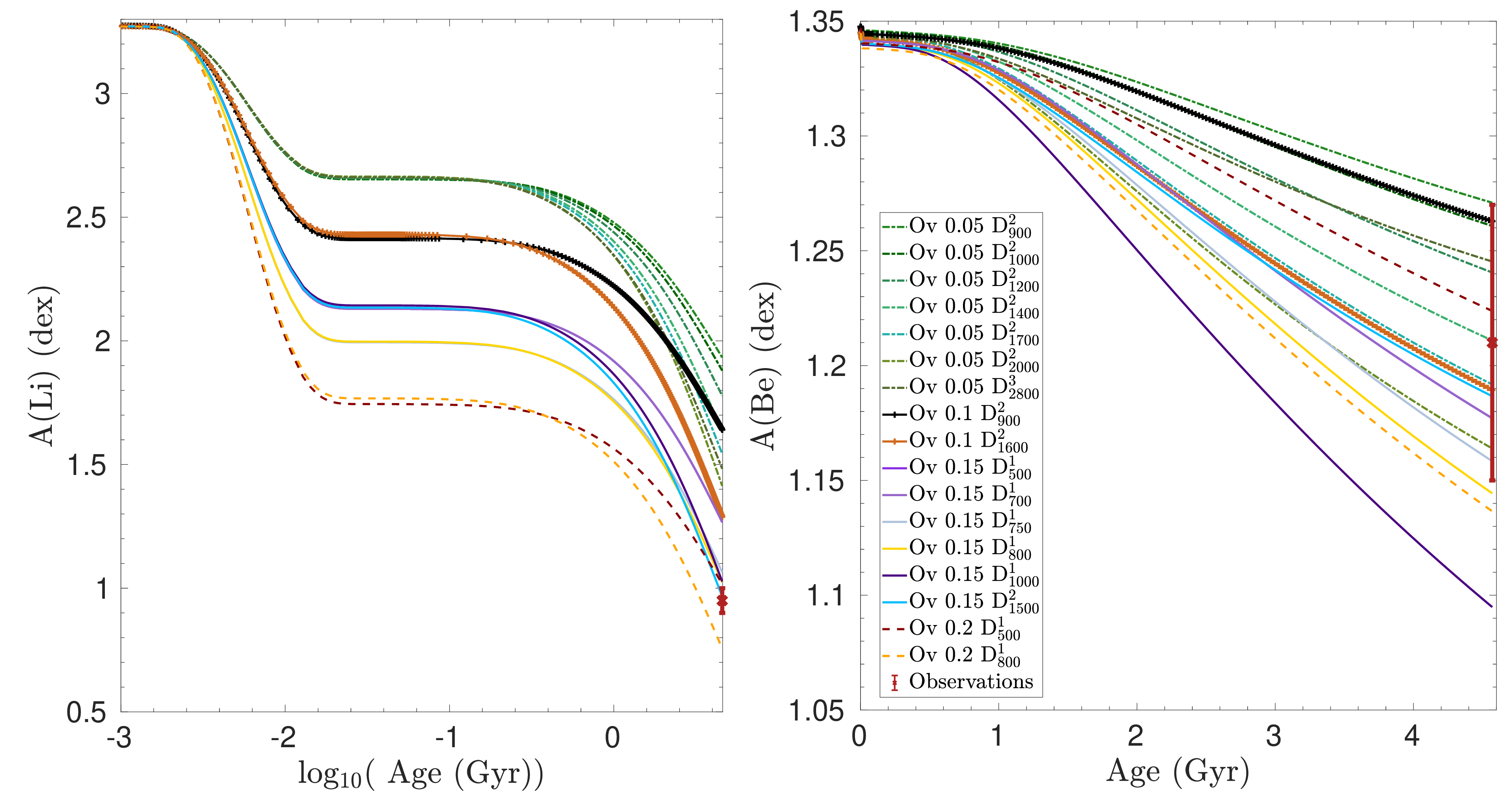}
	\caption{Left panel: Evolution of surface Lithium abundance as a function of age (in log scale) for the models of Table \ref{tabModelsOvDTurb}. The ``Observed'' value is taken from \citet{Wang2021}. Right panel: Evolution of the surface Beryllium abundance as a function of age (in log scale) for the models of Table \ref{tabModelsOvDTurb}. The ``Observed'' value is taken from \citet{Amarsi2024}.}
		\label{Fig:LiBeOVDT}
\end{figure*}

We provide in Table \ref{tabModelsOvDTurb} the fundamental parameters of our models including both overshooting and turbulent diffusive mixing at the BCZ. We illustrate in Figs. \ref{Fig:LiBeOPDT} and \ref{Fig:LiBeOVDT} the lithium and beryllium depletion of our models including overshooting at the BCZ and using the Los Alamos opacities instead of the OPAL opacities. The mixing efficiencies and overshooting values are varied far beyond what would be considered reasonable regarding light element depletion. 

\begin{table*}[h]
\caption{Global parameters of the solar evolutionary models including overshooting and turbulent diffusion.}
\label{tabModelsOvDTurb}
  \centering
\begin{tabular}{r | c | c | c | c | c }
\hline \hline
\textbf{Name}&\textbf{$\left(r/R\right)_{\rm{BCZ}}$}&\textbf{$\mathit{Y}_{\rm{CZ}}$}&\textbf{$\mathit{Y}_{\rm{P}}$}&\textbf{$\mathit{A(Li)}$ (dex)}&\textbf{$\mathit{A(Be)}$ (dex)} \\ \hline
Ov $0.05$ D$^{2}_{900}$&$0.7210$&$0.2461$& $0.2653$ & $1.93$&$1.27$\\
Ov $0.05$ D$^{2}_{1000}$&$0.7210$&$0.2463$& $0.2652$ & $1.88$&$1.26$\\
Ov $0.05$ D$^{2}_{1200}$&$0.7210$&$0.2466$& $0.2650$ & $1.78$&$1.24$\\
Ov $0.05$ D$^{2}_{1500}$&$0.7210$&$0.2470$& $0.2649$ & $1.63$&$1.21$\\
Ov $0.05$ D$^{2}_{1700}$&$0.7211$&$0.2472$& $0.2648$ & $1.54$&$1.19$\\ 
Ov $0.05$ D$^{2}_{2000}$&$0.7212$&$0.2475$& $0.2647$ & $1.41$& $1.16$\\ 
Ov $0.05$ D$^{3}_{2800}$&$0.7211$&$0.2471$& $0.2648$ & $1.48$ & $1.25$\\ 
Ov $0.1$ D$^{2}_{900}$&$0.7166$& $0.2462$ &$0.2652$& $1.64$&$1.26$\\
Ov $0.1$ D$^{2}_{1600}$&$0.7171$& $0.2472$ &$0.2647$& $1.29$&$1.19$\\
Ov $0.15$ D$^{1}_{500}$&$0.7126$&$0.2461$& $0.2653$ & $1.44$ & $1.23$\\
Ov $0.15$ D$^{1}_{700}$&$0.7130$&$0.2469$& $0.2650$ & $1.27$& $1.17$\\
Ov $0.15$ D$^{1}_{1000}$&$0.7135$&$0.2477$& $0.2647$ & $1.03$&$1.09$\\
Ov $0.15$ D$^{2}_{1500}$&$0.7130$&$0.2473$& $0.2647$ & $0.97$& $1.19$\\
Ov $0.17$ D$^{1}_{750}$&$0.7115$&$0.2471$& $0.2649$ & $1.06$&$1.16$\\
Ov $0.17$ D$^{1}_{800}$&$0.7116$&$0.2472$& $0.2649$ & $1.02$&$1.14$\\
Ov $0.2$ D$^{1}_{500}$&$0.7086$&$0.2464$& $0.2652$ & $1.02$&$1.22$\\
Ov $0.2$ D$^{1}_{800}$&$0.7091$&$0.2473$& $0.2648$ & $0.76$ & $1.14$\\
\hline
\end{tabular}
\end{table*}

\end{appendix}

\end{document}